\newif\ifeprint \eprinttrue                                         
\ifeprint                                                           
\PassOptionsToPackage{obeyspaces,spaces,hyphens}{url}
\documentclass[rmp,twocolumn,amsmath,letterpaper,                   
  raggedbottom,showkeys,hyphens]{revtex4}                           
\usepackage{times}                                                  
\usepackage{hyperref}                                               
\usepackage[hyphenbreaks]{breakurl}                                 
\hypersetup{pdftitle=                                               
  {Sensor trajectory estimation by triangulating lidar returns},    
  pdfauthor={C. F. F. Karney and S. Kim}}                           
\usepackage{graphicx}                                               
\newcommand{\doi}[1]{\href                                          
  {https://doi.org/#1}{doi:\discretionary{}{}{}#1}}                 
\let\keyword=\keywords                                              
\newcommand{\authorcontributions}[1]{}                              
\newcommand{\funding}{\section*{Funding}}                           
\else                                                               
\documentclass[remotesensing,article,
  submit,pdftex,moreauthors]{Definitions/mdpi}
\firstpage{1}
\pubvolume{1}
\issuenum{1}
\articlenumber{0}
\pubyear{2022}
\copyrightyear{2022}
\datereceived{}
\dateaccepted{}
\datepublished{}
\hreflink{https://doi.org/} 
\fi                                                         
\usepackage{array}

\ifeprint                                                   
\begin{document}                                            
\title{Sensor trajectory estimation by                      
  triangulating lidar returns}                              
\date{\today}                                               
\author{Charles F. F. Karney}                               
\email{charles.karney@sri.com}                              
\thanks{\\\mbox{\hspace{0.6em}}ORCiD: \href                 
  {https://orcid.org/0000-0002-5006-5836}                   
  {0000-0002-5006-5836}}                                    
\author{Sujeong Kim}                                        
\email{sujeong.kim@sri.com}                                 
\thanks{\\\mbox{\hspace{0.6em}}ORCiD: \href                 
  {https://orcid.org/0000-0002-2641-395X}                   
  {0000-0002-2641-395X}}                                    
\affiliation{SRI International,                             
Princeton, NJ 08543-6449, USA}                              
\else                                                       
\Title{Sensor Trajectory Estimation by Triangulating Lidar Returns}
\TitleCitation{Sensor Trajectory Estimation by Triangulating Lidar Returns}
\newcommand{\orcidauthorA}{0000-0002-5006-5836}
\newcommand{\orcidauthorB}{0000-0002-2641-395X}
\Author{Charles F. F. Karney $^{1,2}$\orcidA{} and Sujeong Kim $^{1,3}$\orcidB{}}
\AuthorNames{Charles Karney and Sujeong Kim}
\AuthorCitation{Karney, C. F. F.; Kim, S.}
\address[1]{%
$^{1}$ \quad SRI International, Princeton, NJ 08543-6449\\
$^{2}$ \quad email: \href{mail:charles.karney@sri.com}{charles.karney@sri.com}\\
$^{3}$ \quad email: \href{mail:sujeong.kim@sri.com}{sujeong.kim@sri.com}}
\fi                                                                    

\ifeprint                                                               
\begin{abstract}                                                        
The paper describes how to recover the sensor trajectory for an aerial  
lidar collect using the data for multiple-return lidar pulses.  This    
work extends the work of \citet{gatziolis19} by performing a            
least-squares fit for multiple pulses simultaneously with a spline fit  
for the sensor trajectory.  The method can be naturally extended to     
incorporate the scan angle of the lidar returns                         
following \citet{hartzell20}.  This allows the pitch and the yaw of     
the sensor to be estimated in addition to its position.                 
\end{abstract}                                                          
\else                                                                   

\abstract{
The paper describes how to recover the sensor trajectory for an aerial
lidar collect using the data for multiple-return lidar pulses.  This
work extends the work of \citet{gatziolis19} by performing a
least-squares fit for multiple pulses simultaneously with a spline fit
for the sensor trajectory.  The method can be naturally extended to
incorporate the scan angle of the lidar returns
following \citet{hartzell20}.  This allows the pitch and the yaw of
the sensor to be estimated in addition to its position.
}

\fi                                                                    
\keyword{airborne laser scanning; trajectory; aircraft position;
  simulation; pulse angle}
\ifeprint\maketitle\else                                               
\begin{document}
\fi                                                                    
\def\figuredir{figures}
\section{Introduction}\label{intro}

Lidar data sets, typically provided in the form of ``{\tt las}'' files
\citep{las}, often do not contain information on the location of the
sensor platform as a function of time.  For data sets which include the
GPS time for each return, it is possible to identify the multiple
returns originating from a given lidar pulse and thus determine its
direction.  By combining the data for multiple pulses emitted in a short
time, it is possible to ``triangulate'' for the position of the sensor.
This idea was proposed by \citet{gatziolis19} who showed how to obtain a
full sensor trajectory.

Here we reformulate this problem with a view to obtaining a more
accurate trajectory.  The trajectory is modeled as a cubic spline fit.
Such a fit independently fits the $x$, $y$, and $z$ components of
$\mathbf R(t)$, the position of the sensor.  The {\it unknowns} in this
model are the parameters specifying the cubic splines.  The {\it knowns}
are the positions (and times) of the multiple returns from individual
lidar pulses.  The optimization problem is then to adjust parameters
specifying the trajectory to minimize the mean squared error between the
returns and a ray drawn from the sensor position to the mean position of
the return.  This is a rather complex nonlinear optimization problem.
Fortunately, it is one that is easily handled by the software library
Ceres Solver \citep{ceres}.

\citet{hartzell20} proposed using just the scan angles of the
lidar returns to compute the lidar trajectory.  This information does
not constrain the trajectory as tightly as the multi-return data.
However the framework offered by Ceres Solver is sufficiently flexible
to allow the scan-angle data to be included in the optimization problem.
This allows the trajectory to be estimated even in cases of sparse
multiple returns.  More importantly it allows the yaw and pitch of
the sensor to be estimated.

\section{Fixed sensor}

In order to introduce the concepts, let us start first by assuming
that the sensor is fixed and emits $n$ multi-return pulses, indexed
by $i \in [1,n]$.  We shall only consider the first and last returns
(ignoring any intermediate returns).  We denote positions of the
returns by
\begin{equation} \label{return}
 \mathbf r_i^\pm = \mathbf r_i \pm d_i \mathbf p_i.
\end{equation}
Here $\mathbf r_i^\pm$ are the positions of first and last returns,
$\mathbf r_i$ is their mean, $\mathbf p_i$ is the unit vector in the
direction from the last to the first return, and $d_i$ is half the
distance between the returns.

\subsection{The reverse method}

The goal now is to determine the position $\mathbf R$ consistent
with these returns.  One approach is to consider the $n$ rays
\begin{equation}
 \mathbf r_i + s_i \mathbf p_i,
\end{equation}
where the distance along the ray is parameterized by $s_i$ and to
solve the $3n$ equations
\begin{equation}
 \mathbf r_i + s_i \mathbf p_i - \mathbf R = \mathbf h_i \approx 0
\end{equation}
for the $3+n$ unknowns $\mathbf R$ and $s_i$.  This is an overdetermined
system for 2 or more pulses and we can then use standard linear algebra
methods to find the solution which minimizes $\sum_i h_i^2$, the
so-called least-squares solution.

This is the approach used by \citet{gatziolis19} who consider just pairs
of pulses $n = 2$.  The problem is that the resulting solution for
$\mathbf R$ is typically not the optimal solution for the trajectory
problem because the system of equations does not involve the return
separation $d_i$ so pulses with closely separated returns are treated
equally to pulses with widely separated returns.  In reality, the latter
returns should be weighted more heavily.

Gatziolis and McGaughey address this problem by selecting an {\it optimal}
pair of returns based on the return separation and the angle between
the pulses.  This is based on a weighting function which needs to be
separately estimated.

We use a simplified version of this linear least-squares problem to
find an initial trajectory for our method described below.  Using the
$z$ component of the residue equations to eliminate $s_i$ from the
system, the equations become
\begin{align*}
\biggl( r_{i,x} - \frac{p_{i,x}}{p_{i,z}} r_{i,z} \biggr) -
\biggl( R_x - \frac{p_{i,x}}{p_{i,z}} R_z \biggr) = h_{i,x} \approx 0, \\
\biggl( r_{i,y} - \frac{p_{i,y}}{p_{i,z}} r_{i,z} \biggr) -
\biggl( R_y - \frac{p_{i,y}}{p_{i,z}} R_z \biggr) = h_{i,y} \approx 0.
\end{align*}
We can write this as the explicit overdetermined linear system
\begin{equation}
\mathsf A \cdot \mathbf R - \mathbf B = \mathbf H \approx 0,
\end{equation}
where $\mathsf A$ is the $2n \times 3$ matrix
\begin{equation}
\mathsf A = \begin{pmatrix}
\vdots & \vdots & \vdots \\
1 & 0 & - p_{i,x}/p_{i,z} \\
0 & 1 & - p_{i,y}/p_{i,z} \\
\vdots & \vdots & \vdots
\end{pmatrix}
\end{equation}
(two rows for each of the $n$ pulses), $\mathbf B$ is the $2n$
column vector
\begin{equation}
\mathbf B = \begin{pmatrix}
\vdots \\
r_{i,x} - (p_{i,x}/p_{i,z}) r_{i,z} \\
r_{i,y} - (p_{i,y}/p_{i,z}) r_{i,z} \\
\vdots
\end{pmatrix},
\end{equation}
and $\mathbf R$ is the unknown sensor position.

This reduces the problem to $2n$ equations for $3$ unknowns.  In this
formulation we determine the horizontal plane $z = R_z$ in which the
rays are most tightly clustered.  This is {\it not} the same problem as
before; however with typical aerial lidar collects the two solutions for
$\mathbf R$ will be reasonably close.  The difference is immaterial in
our application since this solution for $\mathbf R$ is only used as an
initial estimate.

It's also possible to extend this method to allow the position of the
sensor to be a function of time, for example,
\begin{equation*}
  \mathbf R = \mathbf R_0 + \mathbf V t.
\end{equation*}
The least-squares problem is now
\begin{equation}\label{lsprob}
\mathsf A \cdot
\begin{pmatrix}
  \mathbf R_0\\
  \mathbf V
\end{pmatrix}
- \mathbf B = \mathbf H \approx 0,
\end{equation}
where $\mathsf A$ is now the $2n \times 6$ matrix
\begin{equation}
\mathsf A = \begin{pmatrix}
\vdots & \vdots & \vdots & \vdots & \vdots & \vdots \\
1 & 0 & - p_{i,x}/p_{i,z} & t_i & 0   & - t_i p_{i,x}/p_{i,z} \\
0 & 1 & - p_{i,y}/p_{i,z} & 0   & t_i & - t_i p_{i,y}/p_{i,z}\\
\vdots & \vdots & \vdots & \vdots & \vdots & \vdots
\end{pmatrix}
\end{equation}
(and $\mathbf B$ is unchanged).  Here $t_i$ is the time of the $i$th
pulse.

Because pulses with widely separated returns constrain the possible
position of the sensor more strongly than those with nearby returns, we
multiply the rows in $\mathsf A$ and $\mathbf B$ associated with the
$i$th pulse by $d_i$, thereby appropriately {\it weighting} the
least-squares problem.

As a practical matter, Eq.~(\ref{lsprob}) can be solved to give $\mathbf
R_0$ and $\mathbf V$ by a suitable linear algebra package.  For example,
its solution can be obtained using \citet{eigen} with, for example,
\begin{equation*}
\text{\tt RV = A.jacobiSvd().solve(B);}
\end{equation*}

\subsection{The forward method}

We term the above method of estimating $\mathbf R$ described above the
{\it reverse} method, because the rays are traced back from the returns
to the sensor.  An alternative is to trace the rays from $\mathbf R$ to
the midpoint of returns, the {\it forward} method.  Thus each ray is
given by
\begin{equation} \label{pulse}
 \mathbf r_i + s_i \hat{\mathbf q_i},
\end{equation}
where $\mathbf q_i = \mathbf R - \mathbf r_i$ and $\hat{\mathbf q_i}$
is the corresponding unit vector.  The rays now all
intersect at $\mathbf R$ and the optimization problem is to find
$\mathbf R$ and $s_i$ such that Eq.~(\ref{pulse}) is approximately equal
to the position of the first return, $\mathbf r^+$, from
Eq.~(\ref{return}), i.e.,
\begin{equation}
  d_i \mathbf p_i - s_i \hat{\mathbf q_i} = \mathbf e_i \approx 0.
\end{equation}
Again we have $3n$ equations with $3 + n$ unknowns.  However the
quantities that are being minimized, $\mathbf e_i$, is the distance
between the ray and the given positions of the returns.  This method
now naturally gives more weight to widely separated returns and the
solution will similarly be more heavily governed by rays forming
well-conditioned triangles.

Incidentally, the ray from $\mathbf R$ to $\mathbf r_i$ passes equally
close to the first and last returns, so it is only necessary to the
minimize the distance to the first returns.

This system of equations is no longer linear, so it cannot be solved by
linear algebra techniques.  However, it is ideally suited for the Ceres
Solver package.  This finds the least-squares solution for nonlinear
optimization problems.  It also features
\begin{itemize}
\item
  Automatic determination of the Jacobian needed to find the solutions.
  This is achieved by writing the formulas in standard notation but with
  the variables declared to be the C++ type
  ``Jet''.  This combines a quantity and
  its derivative and, through overloaded operators and functions,
  follows all the standard rules of differentiation.
\item
  A robust optimization.  A standard problem of least-squares methods is
  that outliers in the data can skew the solution away from the
  ``right'' one.  Ceres Solver includes a variety of ``loss functions''
  which cause the effect of errors in the equations to fall off past
  some threshold.  For example in this case, the threshold for the loss
  functions might be set to $0.01\,\mathrm{m}$, the typical quantization
  error for positions in a {\tt las} file.
\end{itemize}

\subsection{Simplifying the forward method}

We can simplify the problem by observing that $\mathbf e_i$ is minimized
with $s_i \approx d_i$ and that the resulting $\mathbf e_i$ then spans a
two-dimensional space perpendicular to $\mathbf p_i$.  Thus we can
approximate the error $\mathbf e_i$ by projecting the $\mathbf R -
\mathbf r_i$ onto the plane perpendicular to $\mathbf p_i$ at the first
return.  The first step is to convert to a primed coordinate system with
$\mathbf r_i$ at the origin and with the $\mathbf z'$ axis parallel to
$\mathbf p_i$.  This is achieved by the rotation matrix
\begin{equation}
 \mathsf M_i = \begin{pmatrix}
\displaystyle
\frac{p_{i,x}^2 p_{i,z} + p_{i,y}^2}{p_{i,x}^2+p_{i,y}^2} &
\displaystyle
\frac{-(1 - p_{i,z}) p_{i,x} p_{i,y}}{p_{i,x}^2+p_{i,y}^2} &
-p_{i,x} \\[1ex]
\displaystyle
\frac{-(1 - p_{i,z}) p_{i,x} p_{i,y}}{p_{i,x}^2+p_{i,y}^2} &
\displaystyle
\frac{p_{i,x}^2 + p_{i,y}^2 p_{i,z}}{p_{i,x}^2+p_{i,y}^2} &
-p_{i,y} \\[1ex]
p_{i,x} & p_{i,y} & p_{i,z}
\end{pmatrix}.
\end{equation}
This matrix rotates the coordinate system about the axis $\mathbf z
\times \mathbf p_i$.  Applying this translation and rotation to the
sensor position gives
\begin{equation}
 \mathbf q'_i = \mathsf M_i \cdot \mathbf q_i
\end{equation}
Finally we project $\mathbf q'_i$ onto the plane $z' = d_i$, which gives
\begin{equation}
\mathbf e'_i = \frac{d_i}{q'_{i,z}}
\begin{pmatrix}
q'_{i,x}\\[1ex]
q'_{i,y}
\end{pmatrix}.
\end{equation}
Now the number of unknowns is just $3$, the coordinates of $\mathbf
R$, and the number of equations is $2n$, $\mathbf e_i \approx 0$
for each two-component vector $\mathbf e_i$

Solving this least-squares problem with Ceres Solver entails writing a
C++ class implementing a ``residue block''.  The constructor for the
class takes the {\it knowns} for a particular pulse, i.e., $\mathbf
r_i$, $\mathbf p_i$, and $d_i$, and implements a function object
which accepts the {\it unknowns} $\mathbf R$ as input and returns the
residue $\mathbf e_i$.  This entails merely expressing the equations
above as computer code.  The problem is specified by $n$ such
residue blocks and an initial guess for $\mathbf R$ (obtained, for
example, by the reverse linear least-squares problem).  Ceres Solver
repeatedly invokes the function objects while adjusting $\mathbf R$
to minimize $\sum e_i^2$.  Because of the automatic differentiation
built into Ceres Solver, it can compute the Jacobian for the problem
which says how each component of $\mathbf e_i$ changes as each
component of $\mathbf R$ is varied.  This allows Ceres Solver to
vary $\mathbf R$ in an optimal way in its search for the
least-squares solution.

\section{The trajectory computation}

The discussion above solves for the sensor position at a single instant
of time.  Of course, the sensor position is typically moving and it is
convenient to model the motion as a cubic spline.  One approach would be
to perform a series of fixed sensor calculations, e.g., at
$0.01\,\mathrm s$ intervals including for each calculation 10 pulses
sampled at $0.001\,\mathrm s$ intervals and then to fit a spline to
the resulting positions.

This approach has the drawback that some of the positions may be better
approximated than others and the spline fit should respect this.  This
could be achieved by assigning weights to the various position estimated
and this, essentially, is how Gatziolis and McGaughey addressed this
issue.  However this put another layer of complexity into the problem.

In the spirit of Ceres Solver, it makes more sense to pose the
entire exercise as a single least-squares problem.  Let's start by
describing how to express a cubic spline.

\subsection{The cubic spline}

A cubic spline is a piece-wise cubic polynomial function which in our
application we will use to approximate $\mathbf R(t)$.  Each component
of $\mathbf R(t)$ can be fit independently of the others.  So we only
need to consider a cubic spline for a scalar function $f(t)$ defined
between $T_0$ and $T_K = T_0 + K \,\Delta t$.  The time interval
is divided into $K$ blocks
of duration $\Delta t$, with the $k\text{th}$ block
consisting of the interval $T_k \le t < T_{k+1}$ where $T_k = T_0 + k
\,\Delta t$ and $k \in [0,K)$.  At internal block boundaries, $t = T_k$
for $k \in (0, K)$, we require that $f(t)$, $f'(t)$, and $f''(t)$ be
continuous.

We shall specify the cubic polynomial for the $k\text{th}$ block
by the values of $f(t)$ and $g(t) = \Delta t\,f'(t)$ at the block
boundaries.  It is convenient to introduce a scaled centered time
variable the block $\tau = (t - T_k)/\Delta t - \frac12$.  At the
block boundaries, we have $f = f_k$ and $g = g_k$ at $\tau =
-\frac12$ and $f = f_{k+1}$ and $g = g_{k+1}$ at $\tau =
\frac12$.  It is now a simple matter, e.g., by using the algebra
system, \citet{maxima}, to find the polynomial satisfying the boundary
conditions
\begin{equation}
f(\tau) = a_0 + a_1 \tau + a_2 \tau^2 + a_3 \tau^3,
\end{equation}
where
\begin{align*}
 a_0 &= \textstyle\frac18 (4 f_+ - g_-),\\
 a_1 &= \textstyle\frac14 (6 f_- - g_+),\\
 a_2 &= \textstyle\frac12 g_-,\\
 a_3 &= -2 f_- + g_+,\\
 f_\pm &= f_{k+1} \pm f_k, \\
 g_\pm &= g_{k+1} \pm g_k.
\end{align*}

By specifying the cubic polynomial by its values and derivatives at
the block boundaries, we ensure continuity of $f(t)$ and $f'(t)$.
The jump in the second derivative is given by
\begin{equation}
  \Delta f''(T_k) =
  \frac{6 (f_{k+1}-f_{k-1}) - 2 (g_{k+1}+g_{k-1}) - 8 g_k}
       {\Delta t^2}.
\end{equation}
We will add $\Delta f''(T_k) \approx 0$ to the optimization problem.

In some situations, portions of a lidar collect might consist of ``only
returns''.  These are, of course, not useful of determining the sensor
trajectory by this method.  If the stretch of such returns
spans multiple blocks in the cubic spline, then the spline determination
becomes badly conditioned.  We address this by enforcing an {\it
  additional} constraint on the boundaries between two block with few
multiple returns, namely that the third derivative is continuous.  The
jump in the third derivative is
\begin{equation}
  \Delta f'''(T_k) =
  \frac{4f_k - 2(f_{k+1}+f_{k-1}) + (g_{k+1}-g_{k-1})}
       {\Delta t^3}.
\end{equation}
In order to improve the smoothness of the trajectory, we add a
constraint $\Delta f'''(T_k) \approx 0$ for all block boundaries.
However, the weight for this constraint is very small except when the
boundary separates blocks essentially devoid of multiple returns.

\subsection{The optimization problem}

We are now ready to set up the optimization problem for the entire
trajectory.  The knowns are $t_i$, $\mathbf r_i$, $\mathbf p_i$, and
$d_i$ for $n$ pulses.  These are the same as for the fixed sensor
problem with the addition of the time $t_i$ for each pulse.  The
unknowns are the sensor positions, $\mathbf R(t)$ and velocities,
$\mathbf R'(t)$, at the block boundaries $t = T_k$ for $k \in [0,K]$.

The residue block for Ceres Solver now includes the time $t_i$.  Each
pulse is assigned to a particular time block and the constructor
converts this time to the scaled time $\tau$.  The corresponding
function object now takes the trajectory position and velocities at the
block boundaries as input.  It then uses the stored value of $\tau$ to
evaluate the corresponding cubic polynomials for the 3 components of the
sensor position $\mathbf R(t_i)$.  The calculation then proceeds as in
the fixed-sensor case and returns a two-component vector for the
residue.

There is now also a new type of residue block to enforce the continuity
of the acceleration at the internal block boundaries.  The function
object takes $\mathbf R(T_{k\pm1})$, $\mathbf R'(T_{k\pm1})$, and
$\mathbf R'(T_k)$, and returns the jump in $\mathbf R''(T_k)$.

Overall there are $6(K+1)$ unknowns (the positions and velocities at the
block boundaries).  The number of equations is $2n$ for the pulse
residues, plus $3(K-1)$ for the acceleration jump constraints, plus,
optionally, another $3(K-1)$ for the constraints on the jump in the
third derivatives.

\section{Including the scan angle data}

The method outlined above depends on there being sufficient multiple
returns present in the data.  \citet{hartzell20} suggested using the
{\it scan angle} of the lidar pulse as an alternative method for
triangulating the position of the sensor platform.  This data can be
seamlessly merged into our method allowing the sensor position to be
estimated even in the absence of multiple returns.  This has the added
benefit that the attitude of the sensor platform can be estimated.

The scan angle of the lidar pulse is the angle measured rightwards from
nadir of the lidar pulse as it sweeps left and right either side of the
sensor platform.  In some {\tt las} formats, this is only recorded
to the nearest whole degree.

We start by determining the direction of the laser pulse given the yaw
$\psi$, pitch $\theta$, and roll $\phi$ of the sensor platform and the
scan angle $\alpha$.  The standard coordinate system is $x$ east, $y$
north, and $z$ up.  Given that the sensor starts
in a reference orientation, level and heading due north, the sensor
orientation is found by rotating by $+\phi$ about the $y$ axis, followed
by a rotation $+\theta$ about the $x$ axis, followed by a rotation
$-\psi$ about the $z$ axis.  In the reference orientation, the lidar
pulse is emitted a direction obtained by rotating the downward vector by
$-\alpha_i$ about the $y$ axis (thus positive $\alpha_i$ is to the right
of the sensor path).  Taking account of the attitude of the sensor
platform, the direction of the pulse is
\begin{equation}\label{rots}
\mathsf N(-\psi \hat{\mathbf z}) \cdot
\mathsf N(+\theta \hat{\mathbf x}) \cdot
\mathsf N(-\alpha_i \hat{\mathbf y}) \cdot (-\hat{\mathbf z}),
\end{equation}
where $\mathsf N(\mathbf n)$ is the matrix giving a right-handed
rotation by $\left| n \right |$ about the axis $\hat{\mathbf n}$.  Note
that $\alpha_i$ includes both the roll of the sensor platform $\phi$ and
the deflection of the lidar pulse relative to the sensor platform; so
$\phi$ does not appear here.

Now consider a lidar pulse emitted with the sensor positioned at
$\mathbf R$ and the lidar return recorded at $\mathbf r_i$ with scan
angle $\alpha_i$, so that the ray from the sensor to the return is
$-\mathbf q_i = \mathbf r_i - \mathbf R$.  We now reverse the
order of rotations in Eq.~(\ref{rots}) to put this ray back in a nominal
reference frame for the lidar pulse,
\begin{equation}
  \mathbf q''_i =
  \mathsf N(+\alpha_i \hat{\mathbf y}) \cdot
  \mathsf N(-\theta \hat{\mathbf x}) \cdot
  \mathsf N(+\psi \hat{\mathbf z}) \cdot (-\mathbf q_i).
\end{equation}
We require that $\mathbf q''_i$ be nearly parallel to the downward
direction $-\hat{\mathbf z}$; or, equivalently, that the horizontally
projected 2-vector
\begin{equation}
  \mathbf a_i = \frac1{q''_{i,z}}
  \begin{pmatrix}
    q''_{i,x}\\[1ex]
    q''_{i,y}
  \end{pmatrix}
\end{equation}
be close to zero.

The components of projected vector $\mathbf a_i$ are in the reference
frame of the sensor; thus the $x$ component reflects an error in the
given scan angle $\alpha_i$, while the $y$ component reflects an error
in the unknown pitch $\theta$.  Because the recorded data for $\alpha_i$
often includes the rather large quantization error of $1^\circ$, we
might wish to weight the $y$ component of $a_i$ more heavily.

The conditions $\mathbf a_i \approx 0$ are just other constraints we can
add to our optimization problem for Ceres Solver.

\section{Implementation and testing}

\begin{table*}[tb]
\caption{\label{params} Parameters used in the trajectory estimation code.
The correspondence between these parameters and the notation in the paper
is: $\delta t_r = \mathtt{dtr}$, $\delta t_s = \mathtt{dts}$, $t_r
= \mathtt{multiweight}$, and $w_s = \mathtt{scanweight}$.}
\small
\begin{verbatim}
dtr = 0.005                # sampling interval multi-return (s) (inf = don't)
dts = 0.005                # sampling interval scan-angle (s) (inf = don't)
tblock = 1                 # block size for cubic spline (s)
dr = 0.01                  # error in returns (m)
dang = 1                   # error in scan angle (deg)
scanweightest = 0.01       # relative weight for scan pulses in initial est
multiweight = 5            # weight for multi-return pulses in ceres problem
scanweight = 0.005         # weight for scan pulses in ceres problem
fixedpitch = nan           # fixed pitch value (deg) (nan = don't fix)
pitchweight = 1            # weight pitch vs scan angle
flipscanang = f            # does sign of scan angle need flipping
minsep = 0.5               # min separation (m) of returns considered
accelweight = 0.5          # weighting for acceleration constraint
clampweight = 0.0001       # weighting for general clamping constraint
straddleweight = 0.0001    # weighting for straddling clamping constraint
attaccelweight = 0.01      # weighting for acceleration constraint on attitude
attclampweight = 0.0001    # weighting for clamping constraint on attitude
extrapitchclamp = 1        # extra clamping for pitch component of attitude
vlevel = 0                 # verbosity level
estn = 20                  # min number of pulses to use of rough estimate
niter = 50                 # number of iterations for ceres-solver
# used by the outside
tout = 0.01                # interval (s) for the reported trajectory
\end{verbatim}
\end{table*}
The solution developed above was implemented as a filter of the Point
Data Abstraction Library \citep{pdal,pdal243}.  It has subsequently incorporated
into the \citet{pdalgit} as a possible ``trajectory'' filter.  The input
should be a {\tt las} file for an aerial lidar collect including the GPS
time, return number, and the scan angle.  The filter produces a sensor
trajectory (position and orientation) sampled at regular intervals.
The internal parameters that
govern the operation of the filter together with their default values
and a brief description are given in Table~\ref{params}.

Multiple returns are assumed to originate from the same lidar pulse if
the times match.  The filter combines the multi-return and the
scan-angle constraints as follows: the sequence of pulses is broken into
time intervals $\delta t_r$, resp.~$\delta t_s$, for multi-return,
resp.~scan-angle, constraints and the ``best'' pulse is selected for
each type of constraint.  For multi-return constraints, this is the
pulse with the largest separation between first and last returns.  For
scan-angle constraints, the midpoint of a run of pulses with the same
scan angle is chosen (to minimize the effect of quantizing the scan
angle to a whole degree in {\tt las} files).  In addition, weights,
$w_r$ and $w_s$, can be specified separately for the multi-return and
scan-angle constraints.  In this way it is easy to run with just
multi-angle or just scan-angle constraints or some mixture of
constraints.

Our test data was collected by the National Center for Airborne Laser
Mapping (NCALM) in 2017 over the University of Houston.
The data is given in UTM coordinates (zone 15n) and
height above the ellipsoid.  The direction was approximately westerly
with northing $3\,289.5\,\mathrm{km}$ and easting ranging from
$276.3\,\mathrm{km}$ to $271.5\,\mathrm{km}$, i.e., covering about
$4.8\,\mathrm{km}$ in 70 seconds.
We also had access to the smoothed best estimate of
trajectory, SBET, for this collect.  This serves as the ground truth for
the trajectory.

\begin{figure*}[tb]
\begin{center}
\includegraphics[scale=0.75,angle=0]{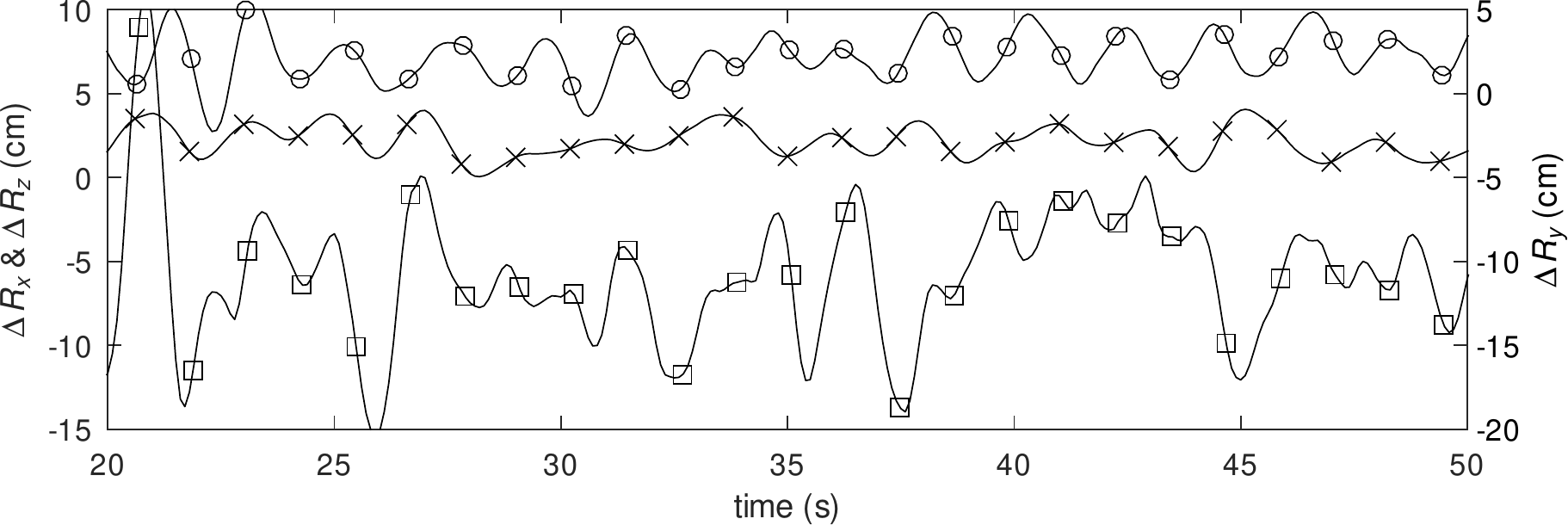}
\end{center}
\caption{\label{ronly}
The discrepancy in the estimated trajectory based only on multi-return
constraints.  The curves give the $x$, $y$, $z$ components of
$\Delta\mathbf R$ marked with crosses, circles, and squares
respectively.  Note well, the scale for $\Delta R_y$ is on the right.}
\end{figure*}

\begin{figure*}[tb]
\begin{center}
\includegraphics[scale=0.75,angle=0]{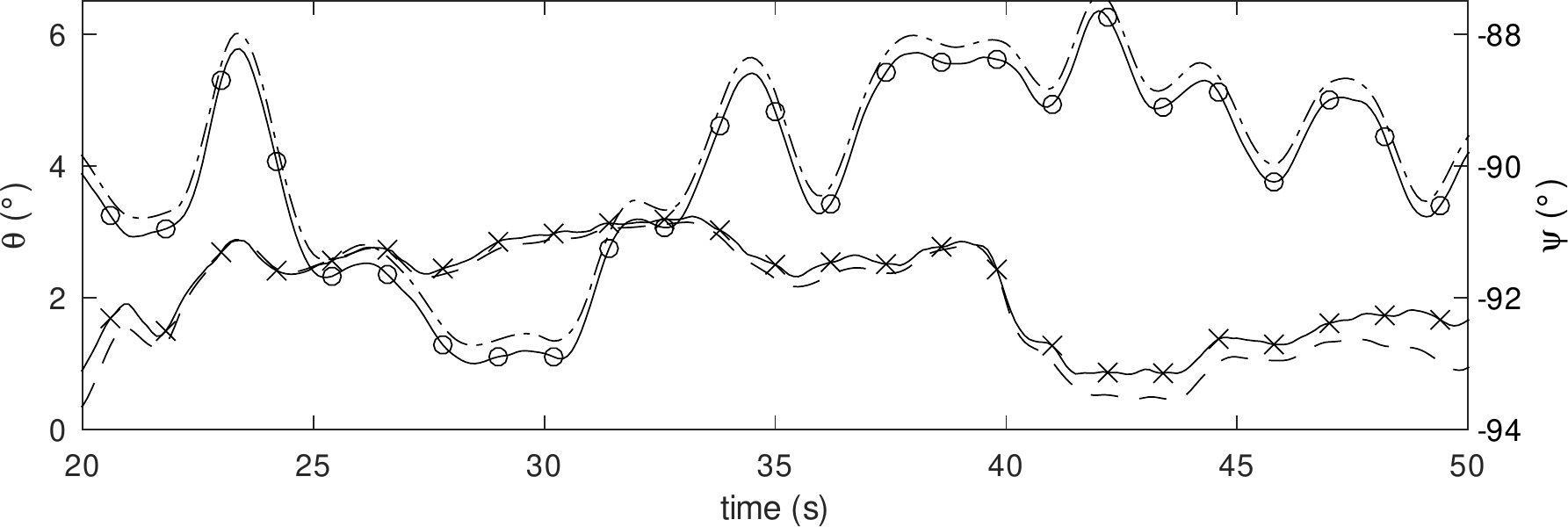}
\end{center}
\caption{\label{sonly}
The pitch and yaw of the sensor using only scan-angle constraints.  The
solid curves show the pitch $\theta$ (marked with crosses) and yaw
$\psi$ (marked with circles).  The dashed and dot-dashed curves show the
corresponding estimates.  Note well, the scale for $\psi$ is on the right}
\end{figure*}

\begin{figure*}[tb]
\begin{center}
\includegraphics[scale=0.75,angle=0]{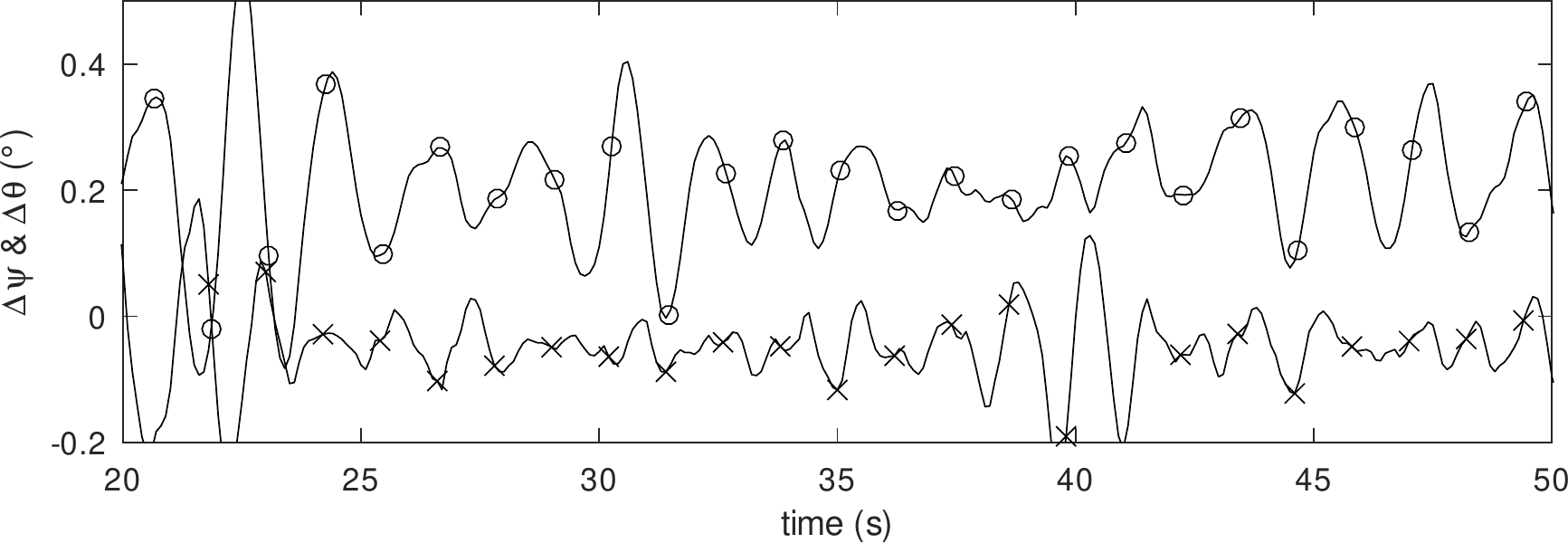}
\end{center}
\caption{\label{rpluss}
The discrepancies in the pitch (marked with crosses) and yaw (marked
with circles) when multi-return and scan-angle constraints are used.}
\end{figure*}

\begin{table}[tb]
\caption{\label{figparams} Parameters used in the figures
and the resulting discrepancies.  Discrepancies are computed over the
full trajectory (of duration $70\,\mathrm s$) and are reported as mean
$\pm$ standard deviation.}
\begin{center}\def\s{\hphantom{-}}
\begin{tabular}{@{\extracolsep{0.5em}}>{$}c<{$}>{$}c<{$}>{$}c<{$}>{$}c<{$}}
\hline\hline\noalign{\smallskip}
& \text{Fig.~\ref{ronly}} & \text{Fig.~\ref{sonly}}
& \text{Fig.~\ref{rpluss}} \\
\noalign{\smallskip}\hline\noalign{\smallskip}
\delta t_r\,(\mathrm{s}) & 0.005 & \infty & 0.005 \\
\delta t_s\,(\mathrm{s}) & \infty & 0.005 & 0.005 \\
w_r & 5 & - & 5 \\ w_s & - & 5 & 0.005 \\
\noalign{\smallskip}\hline\noalign{\smallskip}
\Delta R_x\,(\mathrm m)&\s0.02\pm0.01&\s1.65\pm2.22&
\text{same as Fig.~\ref{ronly}}\\
\Delta R_y\,(\mathrm m)&\s0.02\pm0.01& -1.18\pm0.21&\text{''}\\
\Delta R_z\,(\mathrm m)& -0.05\pm0.05& -4.70\pm0.50&\text{''}\\
\Delta \theta\,({}^\circ)&  - & -0.20\pm0.23& -0.05\pm0.06 \\
\Delta \psi\,({}^\circ)  &  - &\s0.27\pm0.02&\s0.21\pm0.08 \\
\noalign{\smallskip}\hline\hline
\end{tabular}
\end{center}
\end{table}%
We ran the trajectory estimation code with three different settings of
the parameters as described in Table~\ref{figparams}.  Also shown in
this table are the discrepancies in position $\Delta\mathbf R = \mathbf
R_{\mathrm{est}} - \mathbf R_{\mathrm{sbet}}$, in pitch $\Delta\theta
= \theta_{\mathrm{est}} - \theta_{\mathrm{sbet}}$, and in yaw
$\Delta\psi = \psi_{\mathrm{est}} - \psi_{\mathrm{sbet}}$.  The results
are also shown graphically in Figs.~\ref{ronly}--\ref{rpluss}.  We
estimate the trajectory of the entire $70\,\mathrm s$ trajectory;
in these figures, we only display the data for the central
$30\,\mathrm s$ of the flight.

In the first test, we included just multi-return data by setting $\delta
t_s = \infty$; the estimated trajectory is based on the ideas
of \citet{gatziolis19}, albeit using a considerably more sophisticated
optimization strategy.  Even though this involves a system of tens of
thousands of equations, Ceres Solver handles it without difficulty in
about 5 seconds of CPU time.  The results are shown in Fig.~\ref{ronly}
which shows that the estimated trajectory is within a few centimeters of
the ground truth.  On the one hand, this estimate is extraordinarily
good; given that the lidar points are given only to the nearest
centimeter, it's difficult to envision doing substantially better that
this.  On the other hand, perhaps the good fit is unsurprising since,
after all, we are just ``undoing'' the calculation that converted the
raw data from the lidar unit in the points in the {\tt las} file.

The next test, Fig.~\ref{sonly}, switches to using only scan-angle
constraints.  This is an elaboration of the method proposed
by \citet{hartzell20}.  The use of these constraints allow us to
estimate the pitch and yaw of the sensor, in addition to the position.
We see that the estimated values for these components of the orientation
faithfully track the ground truth data.  However the discrepancy in the
position is considerably worse (measured in meters instead of
centimeters) than with using just the multi-return constraints; see the
middle column of Table~\ref{figparams}.  This arises because the
scan-angle constraints are ill-conditioned---a change in pitch can be
compensated by a corresponding shift in position along the line of
travel; note that largest variances in this case are in $\Delta R_x$
(the direction of travel) and $\Delta\theta$.

In the final test, Fig.~\ref{rpluss}, we include both multi-return and
scan-angle constraints.  Because we weight the multi-return constraints
much more heavily, $w_r = 1000 w_s$,
the estimated position is the same as Fig.~\ref{ronly}.  The accurate
position estimate then results in a much better estimate for $\theta$
compared to Fig.~\ref{sonly}.

\section{Discussion}

The method we have described for recovering the sensor trajectory for
lidar data in a {\tt las} file provides accurate results for both the
position and the orientation (yaw and pitch) of the sensor.  It
incorporates the triangulation method using multi-return pulses
of \citet{gatziolis19} and the triangulation method using the scan-angle
of pulses of \citet{hartzell20}.  However in both cases, our method
offers a more robust estimation of the trajectory by performing a single
optimization for the spline fit for the sensor trajectory.  One
important aspect of this optimization is that it automatically accounts
for the conditioning of the triangulation naturally weighting
well-conditioned triangles more heavily.

The example data set we used was for a relatively short flight,
with a dense coverage of returns throughout the flight;
approximately one quarter of the lidar pulse resulting in multiple
returns.  The data for longer flights might need to be broken into
shorter sequences for processing and our method could easily be adapted
to ensure a continuous spline fit throughout the entire flight.  Data
collects with large breaks in the returns, e.g., from flying over water,
would need to be divided at the breaks.  Similarly, we should expect the
performance to degrade with collects giving returns from just one side
of the aircraft, e.g., when flying along a coast, and collects over
terrain resulting in a small fraction of multiple returns.

When discussing the results, we were careful to call the difference
between the estimated trajectory and the ground truth the
``discrepancy''.  This encompasses the irreducible errors because the
{\tt las} data includes quantization errors (typically $1\,\mathrm{cm}$
for position and $1^\circ$ for the scan angle).  In addition, there will
be systematic errors because we treated
UTM coordinates plus height as a Euclidean
coordinate system.  This ignores the fact that $z=0$ is not flat, the
scale of the horizontal dimension is not unity (leading to errors in the
height estimate), and the meridian convergence (leading to errors in the
estimate of the yaw).  However, disentangling these errors from possible
lever arm contributions because of an offset in the inertial navigation
unit and the lidar sensor would require detailed knowledge of the lidar
configuration for our test collect and the post-processing required in
producing the {\tt las} file.  Given that the discrepancies, a few
centimeters for position and a fraction of a degree for orientation, are
within the expected bounds given the errors in {\tt las} data, pursuing
other sources of the discrepancy would require analysis of a broader
range of test data.

\vspace{6pt}

\ifeprint                                           
\section*{Supplementary materials}                  
The following supporting data is provided in the    
{\tt anc} directory in the source package on arXiv: 
original lidar point cloud from NCALM (with the     
intensity field removed and compressed),            
{\tt C2\_L2.laz};                                   
SBET data (trimmed for this collect),               
{\tt sbet\_047\_IGS08-UTM15N-Ellipsoid-trim.txt};   
our trajectory estimate (using the parameters of    
Fig.~\ref{rpluss}),                                 
{\tt C2\_L2-traj3.txt}.                             
\else                                               
\supplementary{The following supporting information can be downloaded at
\linksupplementary{s1}:
original point cloud from NCALM (with the
intensity field removed and compressed),
{\tt C2\_L2.laz};\\
SBET data (trimmed for this collect),
{\tt sbet\_047\_IGS08-UTM15N-Ellipsoid-trim.txt};\\
our trajectory estimate (using the parameters of Fig.~\ref{rpluss}),
{\tt C2\_L2-traj3.txt}.
}
\fi 

\authorcontributions{CK conceived of recasting the multi-return method
  of Gatziolis and McGaughey as a global optimization problem and
  implemented this approach with the Ceres Solver library.
  Simultaneously, SK implemented the scan-angle method of Hartzell as a
  PDAL filter.  CK integrated the scan-angle method into the global
  optimization code and implemented a PDAL filter front end.  SK was
  responsible for testing and debugging.  Both authors have read and
  agreed to the published version of the manuscript.}

\funding{This material is based upon work supported by the United
  States Army Engineer Research and Development Center (ERDC) under
  contract FA2487-21-F-1103. The tests described and the resulting data
  presented herein, unless otherwise noted, are supported under PE
  0602146A ``Network C3I Technology,'' Project AT7 ``Network-Enabled
  Geospatial GEOINT Services Tech'' Task SAT701, 3D Terrain Automated
  Geospatial Co-Registration and Change Detection Algorithms.  Any
  opinions, findings and conclusions or recommendations expressed in
  this material are those of the authors and do not necessarily reflect
  the views of the United States Army.}

\acknowledgments{The authors thank Demetrios Gatziolis for providing
  source code for his algorithms, Ryan Villamil for producing the
  resulting trajectories using the method of \citet{gatziolis19}, and
  Preston Hartzell for supplying his code for the scan-angle method.
  The test data for this study was provided by Craig Glennie and Preston
  Hartzell of the University of Houston.
}

\ifeprint                      
\bibliography{lidar-traj}      
\vfill                         
\else                          

\begin{adjustwidth}{-\extralength}{0cm}
\reftitle{References}
\bibliography{lidar-traj}
\end{adjustwidth}

\fi                            
\end{document}